\documentclass{aa} 
\usepackage{graphicx}        
\usepackage{txfonts}
\usepackage{url}
\usepackage{multicol}
\usepackage{soul}
\usepackage{ulem}
\usepackage{xcolor}
\usepackage{hyperref}
\hypersetup{colorlinks=true,urlcolor=blue,citecolor=blue,linkcolor=blue}

\begin{document}

\title{An overview of HMI off-disk flare observations}

\author{D. Fremstad
          \inst{1}
          \and
          J. C. Guevara Gómez 
          \inst{1,2}
          \and 
          H. Hudson 
          \inst{3,4}
          \and
          J. C. Mart{\'i}nez Oliveros
          \inst{4}
          }

\institute{
{Institute of Theoretical Astrophysics, University of Oslo, Postboks 1029 Blindern, 0315 Oslo, Norway}
\and
{Rosseland Centre for Solar  Physics, University of Oslo, Postboks 1029 Blindern, 0315 Oslo, Norway}
\and
{SUPA School of Physics and Astronomy, University of Glasgow, G12 8QQ, UK}
\and{SSL, University of California, Berkeley CA 92037, USA}
\\
\noindent\email{j.c.g.gomez@astro.uio.no}
}

\date{Received 24 December 2022; Accepted 23 February 2023}

\abstract
{White-light continuum observations of solar flares often have coronal counterparts, including the classical `white-light prominence' (WLP) phenomenon.}
{Coronal emissions by flares, seen in the white-light continuum, have only rarely been reported. 
We seek to use modern data to understand the morphology of WLP events.}
{We have identified a set of 14 examples of WLPs detected by the Heliospheric and Magnetic Imager (HMI) experiment on board the Solar Dynamics Observatory (SDO) satellite using a new online catalogue that covers 2011-2017.
These WLPs invariably accompanied white-light flare (WLF) emission from the lower atmosphere from flares near the limb, as identified by hard X-ray images from the Reuven Ramaty High Energy Spectroscopic Imager (RHESSI).
HMI provides full Stokes information, and we have used the linear polarisations (Q~and~U) to distinguish Thomson scattering from cool material.}
{The event morphologies fit roughly into three categories -- ejection, loop, and spike -- but many events show multiple phenomena.}
{The coronal white-light continuum, observed by HMI analogously to the observations made by a coronagraph, reveals many examples of coronal emission and dynamics.
Using the Stokes linear polarisation, we estimate the masses of hot coronal plasma in 11 of the 14 events and find them to be similar to typical coronal-mass-ejection masses, but without exceeding 10$^{15}$\,g.
We note that the HMI observations do not occult the bright solar disk and were not designed for coronal observations, resulting in relatively low signal-to-noise ratios.
We therefore believe that future such observations with better optimisation will be even more fruitful.
}


\maketitle

\section{Introduction}
     \label{S-Introduction} 
     
The Heliospheric and Magnetic Imager (HMI) has provided a large volume of solar observations at visible wavelengths in a narrow band, including the photospheric Fe~{\sc i} line at 6173.34~\AA\ \citep{2012SoPh..275..207S,2014SoPh..289.3483H}.
These observations include a glimpse of the continuum near this wavelength and have sufficient time resolution to have captured many solar flares.
The observations began on 21 April 2010 and continue to the present, thus including the full sunspot maximum of Solar Cycle 24.
HMI observes solar flares only because it patiently records the full Sun, with a very high duty cycle thanks to the geosynchronous orbit of the Solar Dynamics Observatory \citep[SDO;][]{2012SoPh..275....3P}.
Because HMI's principal scientific objectives are related to helioseismology and  solar magnetism, its flare observations are serendipitous and not optimal; nevertheless, HMI has produced a unique record of the many events that comprise the survey reported here.

Another serendipitous bonus is that HMI images extend into an annular region outside the solar disk, and \cite{2014ApJ...780L..28M} note that two X-class solar flares (SOL2013-05-13T02\footnote{We name the events following the IAU target convention: SOLyyyy-mm-ddThh:mm:ss} and SOL2013-05-13T16) were detectable both on the disk \citep[akin to classical white-light flares; see {e.g.}][]{1989SoPh..121..261N,2016SoPh..291.1273H} and {above the limb}.\ The instrument has thus provided the first systematic database for studying `white-light prominences' (WLPs), observations of which had been extremely rare.
Using HMI's full Stokes capability, \cite{2014ApJ...786L..19S} found linear polarisation in some of these ejecta, making it possible to use the familiar tool of Thomson scattering to explore the flare corona right down to the limb of the Sun.
HMI can thus detect two classes of flare events, which we refer to as on-disk (analogous to white-light flares) and off-disk (the WLPs).
These often go together, especially for events near the limb.
The off-limb component appears in the annular zone (the `annulus'), some tens of arcseconds wide, just above the visible limb.
HMI detects both types of flare events in the `pseudo-continuum' \citep{2018ApJ...860..144S}.
This paper introduces a catalogue of HMI flare observations, the Reuven Ramaty High Energy Spectroscopic Imager (RHESSI) HMI White Light Flare Catalogue, which is available online\footnote{\url{http://sprg.ssl.berkeley.edu/~oliveros/wlf_catalogue/catalog.html}}.
We also review the morphology of the events, which have some novel features.
To our knowledge, there is no other systematic catalogue of HMI off-disk events, but we note that \cite{2017ApJ...851...91N} reported a large survey of on-disk events as part of research on analogous stellar flares.

\section{The catalogue}\label{sec:catalog}

Most white-light flare (WLF) observations have come from broadband sensors, or (rarely) spectroscopic data.
The HMI database we use differs substantially from either of these approaches: the `Ic' pseudo-continuum data product measures the continuum background for a photospheric absorption line over a narrow spectral band, using six wavelength samples over a tuneable range of 678~m\AA\ at the 6173.3~\AA\ line \citep{2014SoPh..289.3483H}.
This limited sampling has some disadvantages; it is a narrow slice of the continuum adjacent to a spectral line that wobbles in wavelength owing to the spacecraft Doppler motion, for example.
In sunspot regions, where WLFs usually happen, the continuum level may be much less than that of the quiet photosphere.
The line itself may actually go into emission during the flare.
Despite these caveats, the flare observations do correlate well
with independent observations as regards image morphology and time variations, with some uncertainties as regards absolute photometry \citep[e.g.][]{2018ApJ...860..144S}.

Our catalogue (at the time of writing) lists 461 entries for M- and X-class flares in Cycle~24 and is based on the RHESSI flare list for the years 2011 through 2016. 
Each entry contains intensity, running-difference, and power-intensity movies for flares within $70$ degrees of disk centre and intensity, plus running-difference and saturated running-difference movies for limb and off-limb events.
A linked script then returns a cutout HMI data cube for any selected event (see the catalogue entry\footnote{Event No.~14 at \url{https://sites.google.com/berkeley.edu/hsi-hmi-catalog/2013}} for the first off-limb detection, of SOL2013-05-13T16:00 (X2.8), as reported by \citealt{2014ApJ...780L..28M}).
This remarkable event is one of the best detections in the HMI annulus, but the catalogue contains many more examples, as surveyed in Sect.~\ref{sec:morph}.

Difference imaging is the essential tool underlying most HMI flare observations.
Occasionally, WLF events on disk are bright enough to be seen without this advantage \citep[famously][]{1859MNRAs..20...13C,1859MNRAs..20...16H}.
In the annular region outside the disk, though, recognition is a little bit harder.
This region is dominated by scattered light originating in the HMI optics, and the noise fluctuations of this unwanted signal dictate the sensitivity limit for flare observations.
The annulus initially was limited to about 20$''$ in width, owing to SDO telemetry limitations. It was increased substantially after the 2013 observations, revealing a number of interesting phenomena in the region.
This improvement can be seen by comparing the examples in Fig.~\ref{fig:morphologies}; it roughly doubled the width of the annulus to about 40$''$.

\section{Morphologies of off-limb white-light flares}\label{sec:morph}

Using the catalogue, we were able to explore the coronal morphology of WLFs systematically. 
This was done by looking for flares with visible mass ejection in the HMI annulus region, where the detection sensitivity is much higher than on the disk because it has no direct photospheric signal. 
In this paper we have manually picked out 13 events from the catalogue in which we can clearly observe activity above the limb. We have also included the SOL2017-09-10T flare, for a total of 14 events.
The continuum in general has several possible contributions \citep{1992PASJ...44...55H,2018ApJ...867..134J}, and, because of the novelty of this database, we make no assumptions about the physical nature of the phenomena, instead
proceeding initially just from morphological appearances.
Candidate physical phenomena would include surges, sprays, jets, filament eruptions, coronal mass ejections, and flare-induced coronal `evaporation', for example.

\begin{figure*}[ht]
    \centering
    \includegraphics[width=\linewidth]{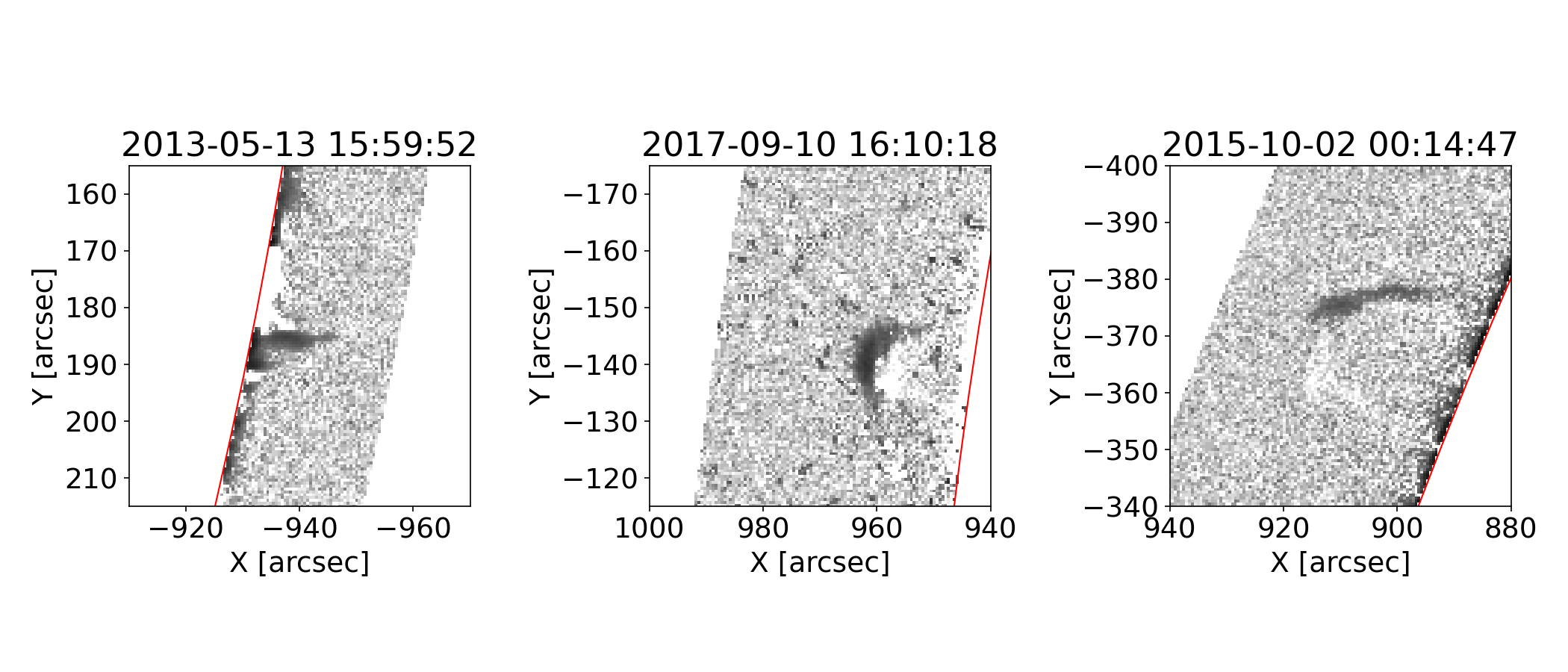}
    \caption{Example of the spike (left; SOL2013-05-13T01:50), ejection (middle; SOL2017-09-10T), and loop (right; SOL2015-10-02T00:07) coronal flare morphologies.
    The date and time of each snapshot is listed above each figure.
    Note the increase in annulus width between 2013 and 2015.
    These are running difference images, from the 45-second cadence of the original data, with black showing newly appearing emission.}
    \label{fig:morphologies}
\end{figure*}

We made rough morphological classifications of the flares by looking at the difference in intensity over time for each flare and observing the dynamics of the flare via movies and difference images, as contained in the catalogue. 
We established three tentative morphologies (see Sect.~\ref{S-morphologies} and Fig.~\ref{fig:morph_cartoon}): spikes, which appear as quick flashes of light close to the limb; ejections -- associated with loop prominence systems (LPSs) in the classical H$\alpha$ sense of \citep{1964ApJ...140..746B} -- which appear as mass moving rapidly through the annulus and are observed to take place after spikes; and loops, which are arc-shaped features appearing to move slowly outwards from the limb. Some events exhibit multiple morphological properties, and typically the spike component precedes the loop component in time, mirroring the impulsive and gradual phases of flare development.
Examples of these morphologies are shown in Fig.~\ref{fig:morphologies}; the images are shown as they come by default in HMI data (i.e. they are not rotated).

\subsection{Analysis of on-disk flare properties} \label{S-inlimb} 

\begin{figure}[ht]
    \centering
    \includegraphics[trim={4cm 0 7cm 0},clip,width=\linewidth]{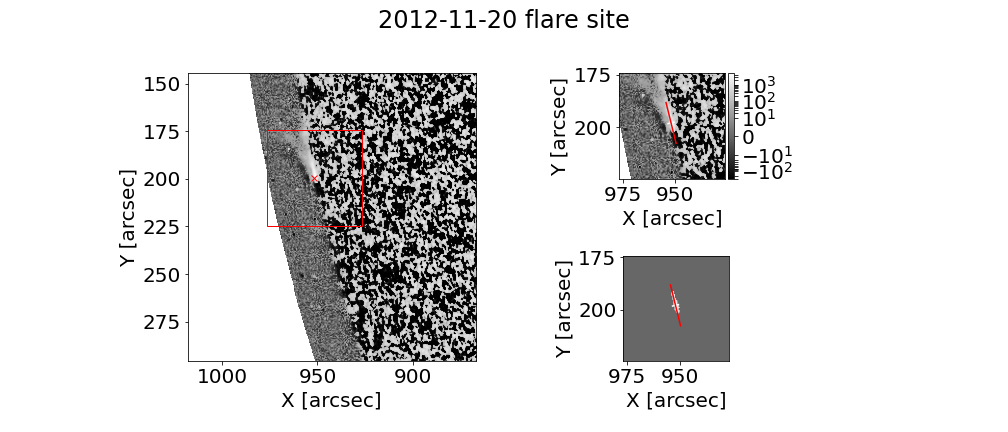}
    \caption{Flare site and kernel fit for SOL2012-11-20T12:27, an ejective event. 
    This is again a running-difference frame, with new emission shown in white. }
    \label{fig:wlb}
\end{figure}

With modern observations it is fairly common to observe some continuum emission from a flare (the WLF phenomenon), especially for the more energetic cases but also sometimes for minor events \citep{2008ApJ...688L.119J}.
In HMI movies, such a flare appears as a quick flash of light on the disk, and, upon closer inspection, they often display a two-ribbon or kernel structure \citep[e.g.][]{2006SoPh..234...79H}, often with diffuse brightening.
Disk events not sufficiently close to the limb cannot be linked to the morphology of the off-disk event, unfortunately. 
The catalogue includes the helio-projective position of the flare, its size, and the time, using the RHESSI onset time for the time field. 
It includes animations of the full data and of difference images, 
selecting the image frame in which the flare is clearly visible. From there we restricted the region and what intensities we included. 
For the classification, we fitted an asymmetric 2D Gaussian kernel to the image frame, which gives us the position, length, and width of the flare brightening:
\begin{equation}
I = I_0 e^{-\big(a(x-x_0)^2 + 2b(x-x_0)(y-y_0) + c (y-y_0)^2\big) },\\ 
\end{equation}
where
\begin{equation}
a = \frac{\cos^2 \theta}{2 \sigma_x^2} + \frac{\sin^2 \theta}{2\sigma_y^2} \\
b = - \frac{\sin 2 \theta}{4 \sigma_x^2} + \frac{\sin 2 \theta}{4 \sigma_y^2} \\ 
c = \frac{\sin^2 \theta}{2 \sigma_x^2} + \frac{\cos^2\theta}{2\sigma_y^2}. 
\end{equation}There are thus six free parameters: the helio-projective position ($[x, y]$), the length and width ($[\sigma_x, \sigma_y]$), the peak intensity ($I_0$), and the rotation angle ($\theta$).
This process was done manually; an example appears in Fig. \ref{fig:wlb}.
In this case, the proximity to the limb of the on-disk emission results in a highly elongated 2D Gaussian fit. Table~\ref{tab:kernels} summarises the kernel parameters for the on-disk events in the catalogue.

\begin{table*}[ht]
\centering
\caption{On-disk flare kernel parameters. 
The first column is the start time of the flare as listed in the RHESSI catalogue.
The second column is the time of the white-light brightening.
The third and fourth columns give the helio-projective $x$ and $y$ coordinates in arcseconds. The fifth and sixth columns describe the standard deviations of the Gaussian kernel along its major ($\sigma_2$) and minor ($\sigma_1$) axes. 
The seventh column shows the centre-averaged raw intensity of the flare with respect to the average intensity of the solar disk centre. The last column shows the rotation angle of the Gaussian kernel with respect to the vertical axis, i.e. the clockwise rotation of the kernel with respect to north-south in degrees. \label{tab:kernels}}
\bigskip

    \begin{tabular}{ |p{4cm}||p{1.cm}|p{1.cm}|p{1.cm}|p{1.cm}|p{1.cm}|p{1.cm}|p{1.cm}|}
    \hline
    Event (RHESSI Start Time) & $t_{WL}$ & $x_0$ & $y_0$ & $\sigma_1$ & $\sigma_2$ & $I_0/I_\odot$ & $\theta$ \\
    \hline
    SOL2011-01-28T00:46 & 00:57 & 936 & 277 & 0.2 & 0.5 & 0.11 & -19 \\
    
    SOL2011-09-22T10:53 & 10:52 & -921 & 159 & 0.4 & 0.5 & 0.35 & 0 \\
    
    SOL2011-10-31T17:48 & 17:47 & -937 & 204 & 0.2 & 0.5 & 0.51 & -2  \\
    
    
    SOL2012-03-02T17:54 & 18:40 & -914 & 305 & 0.5 & 0.9 & 0.41 & 34 \\ 
    
    SOL2012-11-08T02:06 & 02:19 & -926 & 233 & 0.5 & 1.2 & 0.55 & 55  \\ 
    
    SOL2012-11-20T12:28 & 12:39 & 952 & 198 & 0.3 & 2.5 & 0.17 & -14 \\
    
    SOL2013-05-13T01:50 & 02:08 & -930 & 192 & 0.3 & 9 & 0.12 & 11 \\
    
    SOL2013-05-13T15:50 & 15:59 & -931 & 181 & 1.2 & 6.3 & 0.32 & 16 \\
    
    SOL2014-10-16T12:58 & 13:01 & -937 & -221 & 0.4 & 4.5 & 0.19 & -12 \\
    
    SOL2014-11-03T22:08 & 22:31 & -936 & 244 & 0.4 & 9.8 & 0.24 & 14 \\
    
    SOL2015-03-02T15:15 & 15:24 & 899 & 360 & 0.2 & 8 & 0.19 & -20 \\
    
    SOL2015-10-02T00:08 & 00:10 & 819 & -364 & 1.7 & 3.3 & 0.60 & -2 \\
    
    SOL2015-10-02T12:21 & 12:30 & 883 & -330 & 0.3 & 0.9 & 0.61 & 16 \\
    
    
    SOL2017-09-10T16:05 & 15:58 & 942 & -153 & 0.5 & 0.7 & 0.2 & 2 \\
    
    \hline
    \end{tabular}
\end{table*} 

\subsection{Analysis of off-limb flare properties} \label{S-overlimb}

The dynamics seen in the off-limb observations is a key part of the morphological description.
We quantified this by creating time-distance plots, which give visual time lines of the flare dynamics and can be further used to estimate the plane-of-the-sky velocity of the ejected mass. 
These plots show the evolution of ejected mass in height over time and allow us to more easily see the dynamics of the ejected mass. They thus give us a better understanding of the flares' morphology. 
There are several steps that go into making a plot that displays the dynamics of the flare clearly. 
We began by cropping out the solar disk and rotating the frame such that the solar limb is horizontal. 
We selected a central pixel, which preferably is centred on the main body of the ejected mass. 
Next we selected some distance, $d$, from the central pixel and calculated the mean intensity from pixels with a distance less than $d$ from the central pixel. 
This was done for each height in the frame and left us with a strip of mean intensities for different heights. 
We subtracted the continuum intensity from this at each height, which was found by calculating the mean intensity at each height for some time before or after the mass ejection takes place and then taking the mean of these intensities. 
We were thus left with a slice of corrected mean continuum excess intensity. 
Doing this for each frame left us with results such as that shown in the bottom panel of Fig.~\ref{fig:gaussian-fit}. 
We can clearly see how the height of the ejected mass evolves over time, which provides a clear view of the time evolution of the coronal aspects of the flare in the manner that a coronagraphic J-map does \citep{1999JGR...10424739S}.

We calculated speeds for each ejection event by fitting a Gaussian to the signal at each height. The mean of the Gaussian thus traces the flare in time. 
Figure~\ref{fig:gaussian-fit} shows an example of such a trace. 
The mean at each height can then be fitted with a linear function whose slope gives us the velocity projected in the plane of the sky. 
The flare emission has substantial noise fluctuations, resulting in gaps where the Gaussian fits do not 
converge. 
As an example, in Fig.~\ref{fig:gaussian-fit} we use the column of points at the bottom to find the velocity of the first part of the flare (i.e. the spike) and the curve at the top for the second part (the loop), but the points in between them are not relevant and should be ignored. 
We thus made two separate linear fits for the two parts.

\begin{figure}[ht]
    \centering
    \includegraphics[trim={0 0 0 0},clip,width=\linewidth]{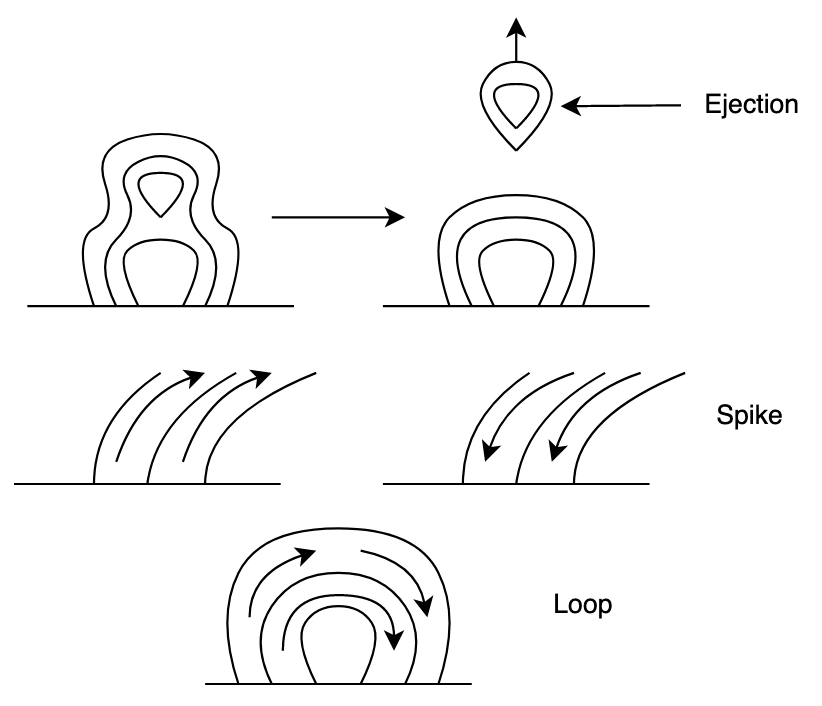}
    \caption{Illustration of the three morphologies. For the LPS or ejection, we see that it is the magnetic field lines themselves that form an independent loop and travel away from the limb. For the spike and loop morphologies, the ejected mass follows the magnetic field lines. In the case of the loop, the ejected mass follows the magnetic field lines all the way around the loop, while for the spike we only observe them to follow it partially.} 
    \label{fig:morph_cartoon}
\end{figure}

\begin{figure}[ht]
    \centering
    \includegraphics[trim={.5cm .5cm .5cm 0},clip,width=\linewidth]{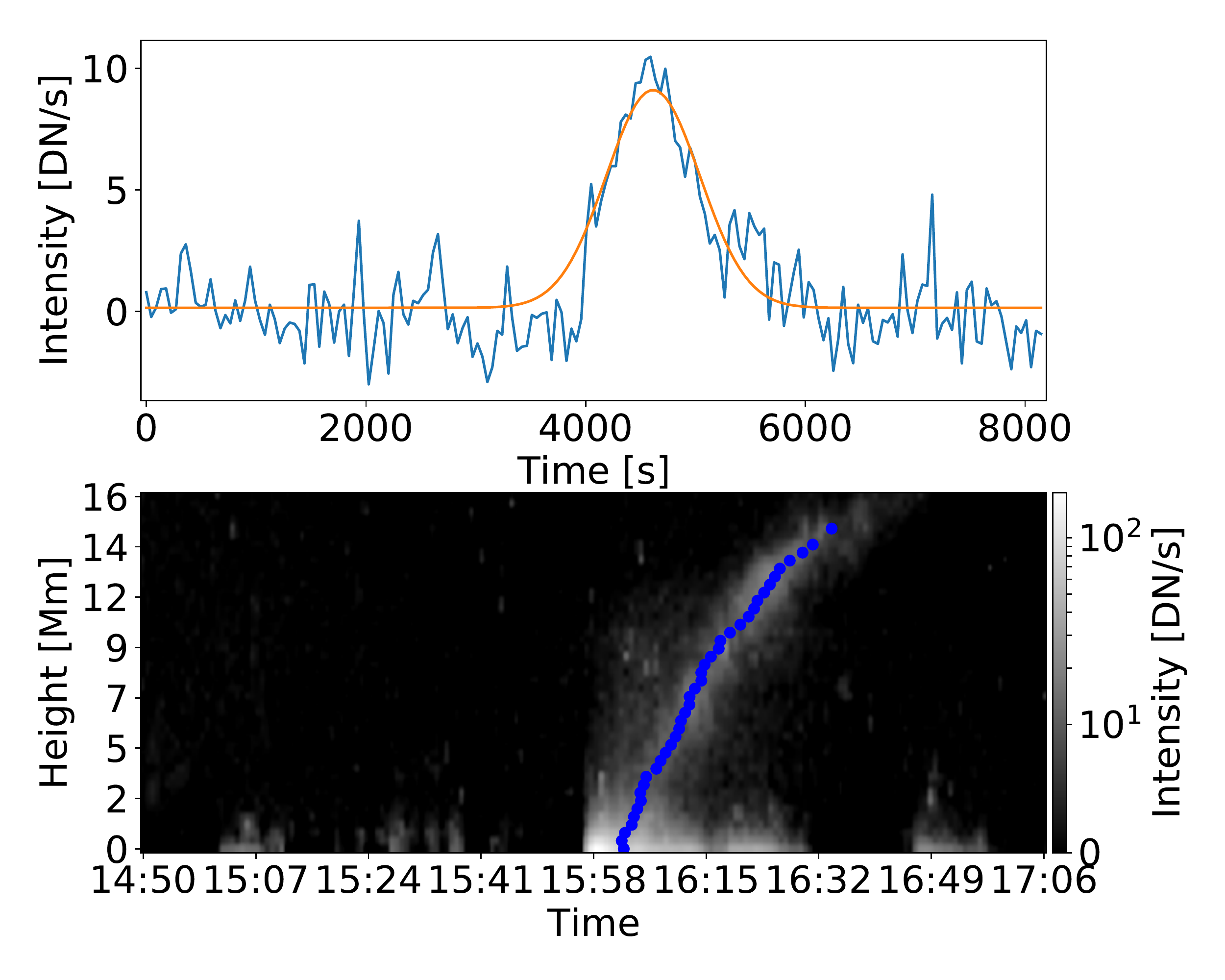}
    \caption{Gaussian fit for flare SOL2013-05-13T15 (middle panel in Fig. \ref{fig:morphologies}). 
    The top panel shows the fit compared to the intensity for a horizontal slice in the middle of the time-distance plot, and the bottom panel shows the trace given by the mean of the Gaussian fit at each height.
    We can see traces of the spike morphology as an illumination at the bottom of the figure, in addition to the ejection (LPS) morphology, which again is seen as a cloud appearing after the spike.}
    \label{fig:gaussian-fit}\label{fig:lpsmorph}
\end{figure}

\subsection{Emission mechanisms} \label{Emission_mech}
\cite{1992PASJ...44...55H} described the mechanisms likely to produce the emissions detected with HMI: Thomson scattering, recombination radiation (free-free and free-bound), and line emission.
These three mechanisms (atomic processes) need not be optically thin if the source densities are large enough \citep{2018ApJ...867..134J}.
HMI's capability for linear polarisation makes detection of the Thomson-scattering component unambiguous, within signal-to-noise limitations.
To disentangle the atomic processes requires consideration of radiative transfer and source geometry. 
\cite{2014ApJ...780L..28M} included detection of the HMI target line (6173~\AA), normally an absorption line in the quiet Sun, which appeared in emission in the legs of the ejection event SOL2013-05-13T16:01 (their Fig.~1).
Generally, we can imagine such sources to be optically thick extensions of the photosphere itself, and thus to have the potential to be much brighter than the Thomson-scattered sources.

\subsection{Masses} \label{S-masses}

We were also able to estimate the event masses and column densities by assuming Thomson scattering as the dominant mechanism.
This is the same as the usual analysis of the K corona \citep{1966gtsc.book.....B}, with the advantage that we can assume a localised source, rather than a spherically extended 3D corona.
We discuss this assumption further below.
The method for calculating these quantities is described in \cite{2022ApJ...936...56M}, where the mass and column density of the SOL2017-09-10 off-limb flare were measured to be of the order of $8 \cdot 10^{14}$ grams and $10^{21}$ cm$^{-2}$, respectively. 

This method requires estimates of the linearly polarised flux, which HMI provides at a 12~min cadence.
The $Q$ and $U$ polarisation components are rotated into the frame of the solar limb such that the rotated $Q'$ component contains all of the linear polarisation for a simple photospheric radiation field \citep[see][]{2021ApJ...923..276S}. 
We defined a region of interest around the Stokes $I$ source and integrated the $Q'$ signal in a pre-flare-subtracted difference image.
We obtained an estimate for each frame to create a time series. 
For some of the events, especially those with a spike morphology, the mass could not be calculated precisely, owing
to the complex and noisy image configuration and its proximity to the limb. 

The mass estimates based on the polarisation reflect only the Thomson scattering and, therefore, may underestimate the total masses of the events. This is because any cool component may not be fully ionised.
We can safely assume that the Thomson component is optically thin, so unless a part of the source is occulted by foreground material, the mass estimate should closely match the amount of material injected at high temperatures into the flare loops \citep[e.g.][]{1972SoPh...23..155H}.
The mass estimation for the spike morphology in particular, or any cool material, would depend upon detailed radiative-transfer modelling, which we have not attempted in this article.

\subsection{Morphologies} \label{S-morphologies}

As previously mentioned, there are three morphologies found in this study; spike, ejection, and loop. With the methods described above we have made time-distance plots and estimated the projected velocities for each of these.
Roughly speaking, spike events seem similar to surges and sprays, ejection events similar to loop prominences, and loop events similar to flows confined to closed coronal fields.
The examples we discuss below show surprises, though, and we cannot be sure that our three morphologies capture all of the possibilities.

Spike morphology is defined by rapid motion, somewhere between 100\,km/s and 300\,km/s. In view of this, the low cadence of HMI image sequences makes polarisation measurements extremely uncertain.
In general, the loss of polarisation data in this way tends to make our mass estimates lower limits.
Spikes can appear as columns of mass ejected away from the solar disk or as large bursts that seemingly do not escape far from the disk and happen very quickly. 
In addition to this, they do not reach very large heights.
We find instances where the ejected mass during spikes moves away from the limb, but also at least one instance with rapid motions towards the limb, which we would associate with coronal rain.
The natural explanation for outward-moving spike events would be as the HMI counterpart of the common H$\alpha$ surge or spray.
An illustration of spike morphology is shown in the middle drawings in Fig.~\ref{fig:morph_cartoon}, and an example of a spike time-distance plot can be found in the bottom panel of Fig.~\ref{fig:gaussian-fit}.
\cite{2014ApJ...780L..28M} also reported spike-like behaviour but with {a downward} motion in SOL2012-05-13 (see below for further comments).
This velocity, plus the appearance of bright material from the corona, strongly suggests an association with the late phase of coronal rain.

Ejection morphology appears to only happen after a spike event. 
It is apparent that the spikes initiate the LPSs and might give rise to the ejected mass that goes into making an LPS. 
The ejection morphology in HMI appears as long clumps of gas that are spread out over a larger period of time in the time-distance plot. 
The low velocity of this morphology causes it to have a very distinct fingerprint on the time-distance plots. 
In Table~\ref{tab:morph} we can see that the velocities of LPSs are usually of the order of magnitude $\sim 10$ km/s, which is consistent with what is found in H$\alpha$ by \citet{1964ApJ...140..746B}.
In this case, the magnetic field forms independent loops that appear to carry the mass away from the limb, as illustrated by the top drawing in Fig. \ref{fig:morph_cartoon}.

\begin{figure}[ht]
    \centering
    \includegraphics[width=\linewidth]{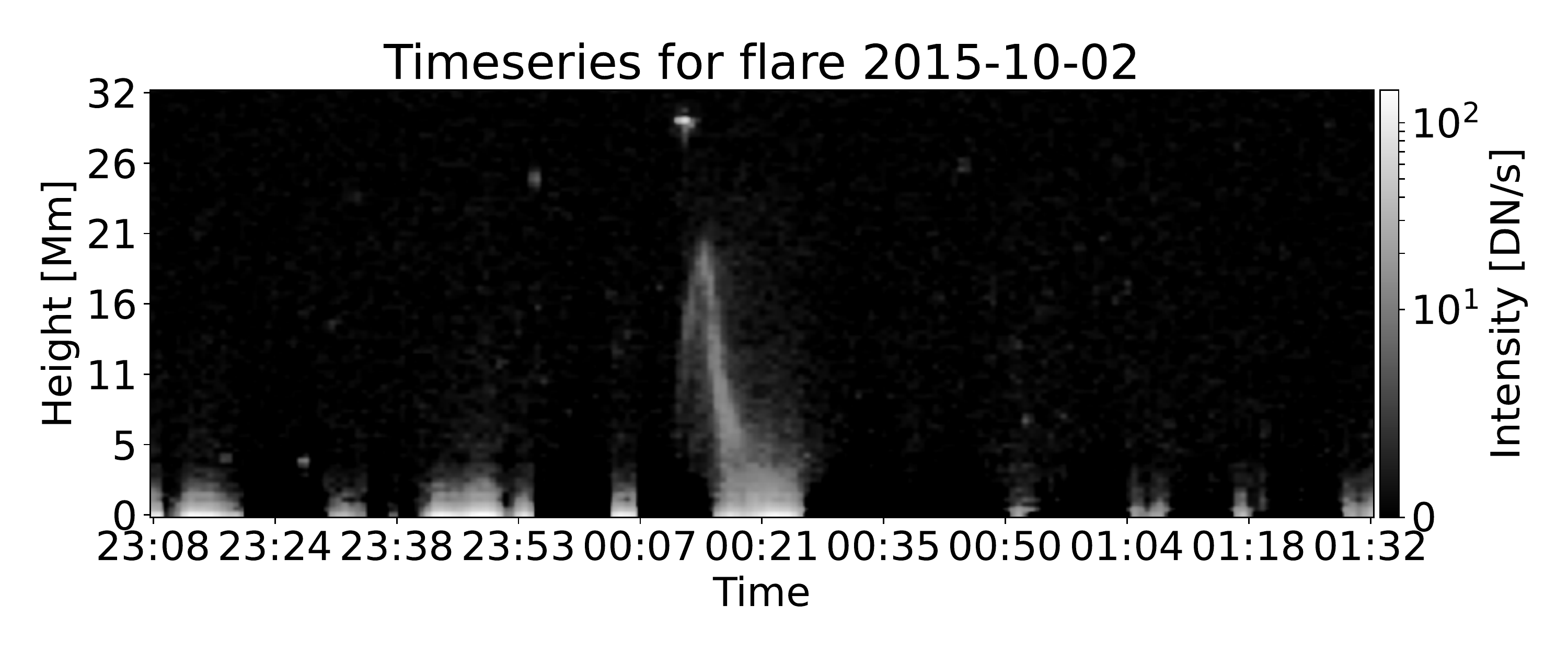}
    \caption{Time series for the SOL2015-10-02 coronal mass ejection, which has a loop morphology.}
    \label{fig:loopmorph}
\end{figure}

Loop morphology appears to reflect parallel flows within loops. Instead of a spike ejecting mass that may eventually form into an LPS, the ejected mass follows an isolated, stationary arch, ascending in one branch and descending in the other \citep{1964ApJ...140..746B}.
Such behaviour suggests the presence of cool material. 
The bottom drawing of Fig. \ref{fig:morph_cartoon} provides an illustration of the loop morphology. 
In the time-distance plot, the loop morphology appears as two long columns connected at the top. 
We can measure the upward and downward velocity of the loop by restricting which part of the time-distance plot we want to fit a Gaussian to. If we cut the plot along the middle of the loop, we can fit a Gaussian and then a linear function to the left side of the loop.\ This gives us the upward-moving velocity, while the right side gives us the velocity moving downwards. 
The loops differ from the LPSs because the former show an apparently parallel flow, do not travel to larger heights, and have higher velocities. 

\begin{table*}[ht]
    \centering
    \begin{tabular}{ |p{4cm}||p{2cm}|p{2cm}|p{2cm}|p{1.2cm}|p{1.2cm}|p{1.2cm}|p{1.5cm}|}
     \hline
     Event & Spike velocity [km/s] & LPS velocity [km/s] & Loop velocity [km/s] & $N_e [cm^{-2}]$ & $m [g]$ & $\Delta Q' / \Delta I$ & $I_0 / I_\odot$\\
     \hline
     SOL2011-01-28T00:46    & -47 $\pm$ 20 & -- & -- & $3 \cdot 10^{20}$ & $10^{14}$ & $0.07$ & $8.1 \cdot 10^{-3}$\\
     
    SOL2011-09-22T10:53    & $61 \pm 35$  & $7 \pm 8$ & -- & $5 \cdot 10^{20}$ & $3 \cdot 10^{14}$ & $0.11$ & $8.6 \cdot 10^{-3}$ \\
    
    SOL2011-10-31T17:47    & --  & $30 \pm 19$ & -- & $4 \cdot 10^{20}$ & $2\cdot 10^{14}$ & $0.02$ & $9.0 \cdot 10^{-3}$ \\ 
     
    SOL2012-03-02T17:54   &   124 $\pm$ 34 &   9 $\pm$ 5  &   -- & -- & -- & -- & -- \\
     
    SOL2012-11-08T02:06    &  -102 $\pm$ 16  & --  & -- & -- & -- & -- & --  \\
     
     SOL2012-11-20T12:28     &   209 $\pm$ 39  &   --  & -- & -- & -- & -- & -- \\
     
     SOL2013-05-13T01:50     &   -       &   18 $\pm$ 10  &   -- & $5 \cdot 10^{20}$ & $3 \cdot 10^{14}$ & $0.02$ & $5.5 \cdot 10^{-3}$ \\
     
     SOL2013-05-13T15:50     &   -     &   15 $\pm$ 5  &   -- & $6 \cdot 10^{20}$ & $3 \cdot 10^{14}$ & $0.02$ & $6.8 \cdot 10^{-3}$\\

     SOL2014-10-16T12:58     &   270 $\pm$ 31  &   --     & -- & $5\cdot 10^{20}$ & $3 \cdot 10^{14}$ & $0.02$ & $6.8\cdot 10^{-3}$ \\
      & -72 $\pm$ 26 & & & & & & \\

     SOL2014-11-03T22:08    &   -64 $\pm$ 11   &   13 $\pm$ 10  &   -- & $2 \cdot 10^{20}$ & $9 \cdot 10^{13}$ & $0.02$ & $6.3 \cdot 10^{-3}$ \\
     
     SOL2015-03-02T15:15     &  -46 $\pm$ 11 & --  & -- & $3\cdot 10^{20}$ & $2 \cdot 10^{14}$ & 0.10 & $6.9 \cdot 10^{-3}$ \\

     SOL2015-10-02T00:08   &   217 $\pm$ 37    &   --   &  -87 $\pm$ 25 & $2\cdot 10^{20}$ & $10^{14}$ & $0.06$ & $7.8 \cdot 10^{-3}$ \\

     SOL2015-10-02T12:21  & -- & 69 $\pm$ 22 & -- & $2\cdot 10^{20}$ & $10^{14}$ & $0.09$ & $8.2 \cdot 10^{-3}$  \\
      
     SOL2017-09-10T16:05 & 33 $\pm$12 & 11 $\pm$ 4 & -- & $2 \cdot 10^{21}$ & $9 \cdot 10^{14}$ & $0.02$ & $7.3 \cdot 10^{-3}$ \\
     \hline
    \end{tabular}
    \smallskip
     \caption{Projected velocities of the respective morphologies. The first column shows the event time. The second, third and fourth columns show the measured (projected) velocity for the morphology type present during that flare. The fifth and sixth columns show the inferred electron 
     density and the total mass of the polarised component, based on peak polarisation signal and avoiding the spike morphology as described in the text. The seventh column shows the measured polarisation, and the eighth column shows the normalised intensity of the ejected material. The large uncertainties in the measurements basically restrict the interpretation to one significant figure.}\label{tab:morph}
\end{table*}

\begin{figure}[ht]
    \centering
    \includegraphics[width=\linewidth]{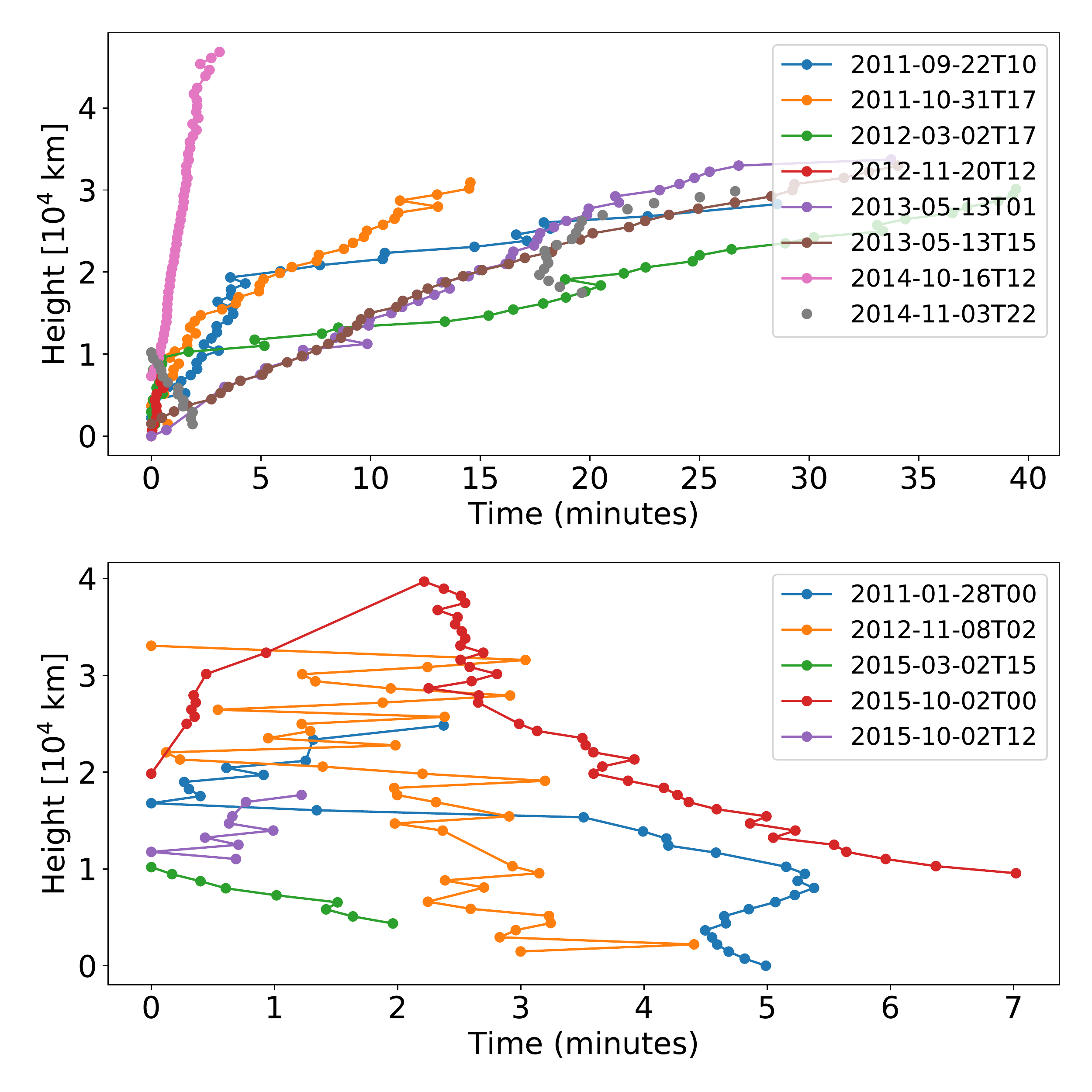}
    \caption{Resulting time evolution in height for the included flares determined by tracing the flares using a Gaussian fit for each height. 
    The flares that are found to be propagating away from the Sun are plotted in the top panel, while those that appear to fall back onto the Sun are plotted in the bottom panel.}
    \label{fig:time_distance_data}
\end{figure}

The complete evolution in height over time for each flare, as determined by tracing the flares by fitting a Gaussian to each height, is plotted in Fig. \ref{fig:time_distance_data}. 
We find that for all cases of LPSs the ejected mass is always propagating away from the Sun. 
Similar results to what is shown in the left panel are also reported by \cite{1964ApJ...140..746B}.
In the case of the spikes, we can also find cases where the mass is moving away from the Sun, but we also find evidence that they can fall down back onto the Sun. 
In addition, the SOL2015-10-02 event  shows evidence of loop morphology, and it can be seen to move away from the Sun before reversing.

Summarising the appearance and behaviour of the HMI off-limb observations, we suggest that the spike morphology corresponds to flare surge or spray activity during the impulsive phase, and that ejection may correspond to filament eruptions.
The ejection morphology is well understood from the classic behaviour of the H$\alpha$ LPSs.
A possible fourth morphology, rain, may be present.
This would describe the infall observed by \cite{2014ApJ...780L..28M} in the event SOL2013-05-15T16:01, in which material appeared suddenly within the annular region and moved downwards at a projected speed estimated at 134$\pm$8\,km/s; the act of appearance strongly suggests an association with thermal instability \citep{1965ApJ...142..531F} and the enhancement of HMI low-temperature opacity, rather than Thomson scattering by hot electrons.
This observation required special analysis at HMI's highest time resolution, using all of the polarisation and wavelength images, each differenced against its preceding like exposure (see \citealt{2017ApJ...839...67S} for a discussion of high-cadence data from HMI).
This kind of analysis was beyond the scope of the present paper, and so this event does not appear in the lower panel of Fig.~\ref{fig:time_distance_data}.

\section{Conclusion} \label{S-Conclusion} 

The pseudo-continuum observations of the Sun from HMI constitute a unique database for characterising solar WLFs, and include the first-ever systematic view of WLF manifestations in the low corona from an observatory in space.
These off-limb observations follow and extend a very limited historical set of data, mainly visual, of WLPs.
This paper introduces an online catalogue\footnote{\url{http://sprg.ssl.berkeley.edu/~oliveros/wlf\_catalogue/catalog.html}} 
with comprehensive coverage of the off-limb events observed by HMI, and further analysis of the on-disk morphology as well.
The catalogue reveals three distinct non-exclusive morphologies -- spike, ejection, and loop -- which produce an effect in an annulus about 40$''$ in width above the visible disk.

The full Stokes capability of HMI shows that Thomson scattering plays a major role in WLF ejecta \citep{2012SoPh..275..207S}.
This is the classic coronagraphic K-corona observational mode, extended to the base of the corona, where the incident radiation field may have significant anisotropy \citep{2021ApJ...923..276S}.
In some cases, the off-limb structures achieve sufficient density (above about $10^{12}$~cm$^{-3}$) to allow normal collisional excitation to play a role \citep{2018ApJ...867..134J}.
The linear polarisation, however, allows for a direct determination of the ejected mass \citep{2022ApJ...936...56M}.

\section*{Acknowledgments}
J.C.G.G is supported by the SolarALMA project, which received funding from the European Research Council (ERC) under the European Union’s Horizon 2020 research and innovation programme (grant agreement No. 682462), and by the Research Council of Norway through its Centres of Excellence scheme, project number 262622. H.H. thanks the School of Physics and Astronomy, University of Glasgow, for hospitality. 

\noindent\textit{Data sources:} All data used in this article can be found in public-domain archives.


\bibliographystyle{aa}
\bibliography{my.bib}

\begin{thebibliography}{23}
\expandafter\ifx\csname natexlab\endcsname\relax\def\natexlab#1{#1}\fi

\bibitem[{{Billings}(1966)}]{1966gtsc.book.....B}
{Billings}, D.~E. 1966, {A guide to the solar corona}

\bibitem[{{Bruzek}(1964)}]{1964ApJ...140..746B}
{Bruzek}, A. 1964, \apj, 140, 746

\bibitem[{{Carrington}(1859)}]{1859MNRAs..20...13C}
{Carrington}, R.~C. 1859, \mnras, 20, 13

\bibitem[{{Field}(1965)}]{1965ApJ...142..531F}
{Field}, G.~B. 1965, \apj, 142, 531

\bibitem[{{Hiei} {et~al.}(1992){Hiei}, {Nakagomi}, \&
  {Takuma}}]{1992PASJ...44...55H}
{Hiei}, E., {Nakagomi}, Y., \& {Takuma}, H. 1992, \pasj, 44, 55

\bibitem[{{Hodgson}(1859)}]{1859MNRAs..20...16H}
{Hodgson}, R. 1859, \mnras, 20, 15

\bibitem[{{Hoeksema} {et~al.}(2014){Hoeksema}, {Liu}, {Hayashi}, {Sun},
  {Schou}, {Couvidat}, {Norton}, {Bobra}, {Centeno}, {Leka}, {Barnes}, \&
  {Turmon}}]{2014SoPh..289.3483H}
{Hoeksema}, J.~T., {Liu}, Y., {Hayashi}, K., {et~al.} 2014, \solphys, 289, 3483

\bibitem[{{Hudson}(2016)}]{2016SoPh..291.1273H}
{Hudson}, H.~S. 2016, \solphys, 291, 1273

\bibitem[{{Hudson} \& {Ohki}(1972)}]{1972SoPh...23..155H}
{Hudson}, H.~S. \& {Ohki}, K. 1972, \solphys, 23, 155

\bibitem[{{Hudson} {et~al.}(2006){Hudson}, {Wolfson}, \&
  {Metcalf}}]{2006SoPh..234...79H}
{Hudson}, H.~S., {Wolfson}, C.~J., \& {Metcalf}, T.~R. 2006, \solphys, 234, 79

\bibitem[{{Jej\-{\v c}i{\v c}} {et~al.}(2018){Jej\-{\v c}i{\v c}}, {Kleint}, \&
  {Heinzel}}]{2018ApJ...867..134J}
{Jej\-{\v c}i{\v c}}, S., {Kleint}, L., \& {Heinzel}, P. 2018, \apj, 867, 134

\bibitem[{{Jess} {et~al.}(2008){Jess}, {Mathioudakis}, {Crockett}, \&
  {Keenan}}]{2008ApJ...688L.119J}
{Jess}, D.~B., {Mathioudakis}, M., {Crockett}, P.~J., \& {Keenan}, F.~P. 2008,
  \apjl, 688, L119

\bibitem[{{Mart{\'\i}nez Oliveros} {et~al.}(2022){Mart{\'\i}nez Oliveros},
  {Guevara G{\'o}mez}, {Saint-Hilaire}, {Hudson}, \&
  {Krucker}}]{2022ApJ...936...56M}
{Mart{\'\i}nez Oliveros}, J.~C., {Guevara G{\'o}mez}, J.~C., {Saint-Hilaire},
  P., {Hudson}, H., \& {Krucker}, S. 2022, \apj, 936, 56

\bibitem[{{Mart{\'{\i}}nez Oliveros} {et~al.}(2014){Mart{\'{\i}}nez Oliveros},
  {Krucker}, {Hudson}, {Saint-Hilaire}, {Bain}, {Lindsey}, {Bogart},
  {Couvidat}, {Scherrer}, \& {Schou}}]{2014ApJ...780L..28M}
{Mart{\'{\i}}nez Oliveros}, J.-C., {Krucker}, S., {Hudson}, H.~S., {et~al.}
  2014, \apjl, 780, L28

\bibitem[{{Namekata} {et~al.}(2017){Namekata}, {Sakaue}, {Watanabe}, {Asai},
  {Maehara}, {Notsu}, {Notsu}, {Honda}, {Ishii}, {Ikuta}, {Nogami}, \&
  {Shibata}}]{2017ApJ...851...91N}
{Namekata}, K., {Sakaue}, T., {Watanabe}, K., {et~al.} 2017, \apj, 851, 91

\bibitem[{{Neidig}(1989)}]{1989SoPh..121..261N}
{Neidig}, D.~F. 1989, \solphys, 121, 261

\bibitem[{{Pesnell} {et~al.}(2012){Pesnell}, {Thompson}, \&
  {Chamberlin}}]{2012SoPh..275....3P}
{Pesnell}, W.~D., {Thompson}, B.~J., \& {Chamberlin}, P.~C. 2012, \solphys,
  275, 3

\bibitem[{{Saint-Hilaire} {et~al.}(2021){Saint-Hilaire}, {Mart{\'\i}nez
  Oliveros}, \& {Hudson}}]{2021ApJ...923..276S}
{Saint-Hilaire}, P., {Mart{\'\i}nez Oliveros}, J.~C., \& {Hudson}, H.~S. 2021,
  \apj, 923, 276

\bibitem[{{Saint-Hilaire} {et~al.}(2014){Saint-Hilaire}, {Schou},
  {Mart{\'\i}nez Oliveros}, {Hudson}, {Krucker}, {Bain}, \&
  {Couvidat}}]{2014ApJ...786L..19S}
{Saint-Hilaire}, P., {Schou}, J., {Mart{\'\i}nez Oliveros}, J.-C., {et~al.}
  2014, \apjl, 786, L19

\bibitem[{{Scherrer} {et~al.}(2012){Scherrer}, {Schou}, {Bush}, {Kosovichev},
  {Bogart}, {Hoeksema}, {Liu}, {Duvall}, {Zhao}, {Title}, {Schrijver},
  {Tarbell}, \& {Tomczyk}}]{2012SoPh..275..207S}
{Scherrer}, P.~H., {Schou}, J., {Bush}, R.~I., {et~al.} 2012, \solphys, 275,
  207

\bibitem[{{Sheeley} {et~al.}(1999){Sheeley}, {Walters}, {Wang}, \&
  {Howard}}]{1999JGR...10424739S}
{Sheeley}, N.~R., {Walters}, J.~H., {Wang}, Y.-M., \& {Howard}, R.~A. 1999,
  \jgr, 104, 24739

\bibitem[{{Sun} {et~al.}(2017){Sun}, {Hoeksema}, {Liu}, {Kazachenko}, \&
  {Chen}}]{2017ApJ...839...67S}
{Sun}, X., {Hoeksema}, J.~T., {Liu}, Y., {Kazachenko}, M., \& {Chen}, R. 2017,
  \apj, 839, 67

\bibitem[{{{\v{S}}vanda} {et~al.}(2018){{\v{S}}vanda}, {Jur{\v{c}}{\'a}k},
  {Ka{\v{s}}parov{\'a}}, \& {Kleint}}]{2018ApJ...860..144S}
{{\v{S}}vanda}, M., {Jur{\v{c}}{\'a}k}, J., {Ka{\v{s}}parov{\'a}}, J., \&
  {Kleint}, L. 2018, \apj, 860, 144

\end{thebibliography}

\end{document}